\documentclass[11pt]{article}
\usepackage{pslatex}
\usepackage{fancyhdr}
\usepackage{graphicx}
\usepackage{geometry}

\RequirePackage{amsfonts,amssymb,amsmath,amscd,amsthm}
\RequirePackage{txfonts}
\RequirePackage{graphicx}
\RequirePackage{xcolor}
\RequirePackage{geometry}
\RequirePackage{enumerate}

\def\figurename{Figure} 
\makeatletter
\renewcommand{\fnum@figure}[1]{\figurename~\thefigure.}
\makeatother

\def\tablename{Table} 
\makeatletter
\renewcommand{\fnum@table}[1]{\tablename~\thetable.}
\makeatother

\usepackage{amsmath}
\usepackage{amssymb}
\usepackage{amsfonts}
\usepackage{amsthm,amscd}

\newtheorem{theorem}{Theorem}[section]
\newtheorem{lemma}[theorem]{Lemma}

\newtheorem{proposition}[theorem]{Proposition}
\theoremstyle{definition}
\newtheorem{definition}[theorem]{Definition}

\theoremstyle{remark}

\numberwithin{equation}{section}

\def\cal{\mathcal}

\begin{document}
\title{\bfseries\scshape{$(q;l,\lambda)$-deformed Heisenberg algebra: coherent states, their statistics and geometry}}
\author{\bfseries\scshape Joseph D\'esir\'e Bukweli-Kyemba\thanks{E-mail address: \tt{desbuk@gmail.com}}\\
University of Abomey-Calavi\\
International Chair in Mathematical Physics\\ and  Applications (ICMPA - UNESCO Chair)\\ 072 B.P.: 50
Cotonou, Republic of Benin\\
\\\bfseries\scshape Mahouton Norbert Hounkonnou\thanks{E-mail address: \tt{norbert.hounkonnou@cipma.uac.bj or hounkonnou@yahoo.fr}}\\
University of Abomey-Calavi\\
International Chair in Mathematical Physics\\ and  Applications (ICMPA - UNESCO Chair)\\ 072 B.P.: 50
Cotonou, Republic of Benin}

\date{}
\maketitle \thispagestyle{empty} \setcounter{page}{1}


\begin{abstract}
The Heisenberg algebra is deformed with the set of parameters
 $\{q, l,\lambda\}$  to generate  a new family of generalized 
coherent states respecting  the Klauder criteria. In this framework, 
 the matrix elements of  relevant operators are exactly computed. Then, 
 a proof on the sub-Poissonian character of the statistics of  the main deformed states 
 is provided. This property   is used to   determine   the induced generalized metric. 
\end{abstract}


\vspace{.08in} \noindent \textbf{Keywords}: Heisenberg algebra, coherent states, Fubini-Study metric, Poisson density.

\section{Introduction}
The Heisenberg algebra is generated by the the identity operator $\mathbf{1}$ and two mutually adjoint operators,  $b$ and   its Hermitian conjugate $b^\dagger$ (also  called annihilation and  creation operators in Physics literature),   satisfying the commutation relations
\begin{eqnarray}\label{Harmonic}
 [b,\;b^\dagger]=\mathbf{1},\qquad [b,\;\mathbf{1}]=0=[b^\dagger,\;\mathbf{1}],
\end{eqnarray}
where  $[A, \;B]= AB - BA.$
Defining the operator  $N:=b^\dagger b,$  known as  the  {\it number operator}, the commutation relations (\ref{Harmonic})
induce the two following  properties:
\begin{eqnarray}\label{Harmonic2}
 [N,\;b]=-b\quad\mbox{and}\quad [N,\;b^\dagger]=b^\dagger.
\end{eqnarray}

Let $\mathcal{F}$  be a Fock space and $\{|n\rangle\;|\;n\in \mathbb{N}\bigcup \{0\}\}$ be its orthonormal  basis. The actions of $b$, $b^\dagger$ and $N$ on  $\mathcal{F}$ are given by 
\begin{eqnarray}\label{Harmonic2b}
 b|n\rangle= \sqrt{n}|n-1\rangle,\;\;b^\dagger|n\rangle=\sqrt{n+1}|n+1\rangle,\;\mbox{ and }\; N|n\rangle=n|n\rangle
\end{eqnarray}
where  $|0\rangle$ is a normalized vacuum:
\begin{eqnarray}
 b|0\rangle= 0,\qquad \langle 0|0\rangle=1.
\end{eqnarray}
From (\ref{Harmonic2b}) the states $|n\rangle$ for $n\ge1$ are built as follows:
\begin{eqnarray}
 |n\rangle=\frac{1}{\sqrt{n!}}(b^\dagger)^n|0\rangle,\;\; n= 1,\;2,\;\cdots
\end{eqnarray}
satisfying the orthogonality and completeness conditions:
\begin{eqnarray}
 \langle m|n\rangle=\delta_{m,n}, \;\quad \sum_{n=0}^\infty |n\rangle \langle n|= \mathbf{1}.
\end{eqnarray}

\pagestyle{fancy} \fancyhead{} \fancyhead[EC]{J. D. Bukweli-Kyemba and M. N. Hounkonnou}
 \fancyhead[EL,OR]{\thepage} \fancyhead[OC]
\fancyfoot{}
\renewcommand\headrulewidth{0.5pt}

\begin{definition}\label{CSdef1}
The normalized states $|z\rangle \in \mathcal{F}$ for $z \in \mathbb{C}$ satisfying one of the  following three equivalent conditions:
\begin{itemize}
\item[(i)] 
\begin{eqnarray}
 b|z\rangle= z|z\rangle,\qquad \langle z|z\rangle=1
\end{eqnarray}
or
\item[(ii)] 
\begin{eqnarray}
 (\Delta Q)(\Delta P)= \frac{\hbar}{2}
\end{eqnarray}
where { $(\Delta X)^2:=\langle z|X^2-\langle X\rangle^2|z\rangle$}  with { $\langle X\rangle:= \langle z|X|z\rangle$},
\begin{eqnarray}
 Q:= \left({\hbar}/{2\mathbf{ m}\omega}\right)^{1/2}(b+b^\dagger),\qquad
P:= -i\left({\mathbf{ m}\hbar\omega}/{2}\right)^{1/2}(b-b^\dagger)\nonumber
\end{eqnarray}
or
\item[(iii)] 
\begin{eqnarray}\label{bch2}
 |z\rangle=e^{zb^\dagger-\bar z b}|0\rangle
\end{eqnarray}
\end{itemize}
are called the coherent states (CS).
\end{definition}
In the condition $(ii),$  $\mathbf{ m}$ stands for the particle mass; $\omega$ is the angular frequency.
Explicitly, the canonical CS are computed as follows:
\begin{eqnarray}\label{bch1}
 |z\rangle=e^{-|z|^2/2}\sum_{n=0}^\infty\frac{z^n}{\sqrt{n!}}|n\rangle= e^{-|z|^2/2}e^{zb^\dagger}|0\rangle,\; z\in\mathbb{C}.
\end{eqnarray}
In (\ref{bch1}) and (\ref{bch2}) we use the famous elementary Baker-Campbell-Hausdorff formula
\begin{eqnarray}
e^{A+B}=e^{-{{1}\over{2}}[A,B]}e^Ae^B
\end{eqnarray}
whenever $[A,[A,B]]=[B,[A, B]]= 0$.
The important feature of these coherent states resides in the partition (resolution) of unity:
\begin{eqnarray}
\int_\mathbb{C}{{[d^2z]}\over\pi} |z\rangle\langle z|= \sum_{n=0}^\infty  |n\rangle \langle n|= \mathbf{1},
\end{eqnarray}
where we have put $[d^2z]= d(Rez)d(Imz)$ for simplicity.
\begin{definition}
The unitary operator
\begin{eqnarray}
D(z):=e^{zb^\dagger-\bar z b},\qquad z\in \mathbb{C}
\end{eqnarray}
is called a coherent (displacement) operator.
\end{definition}
From the property
\begin{eqnarray}
D(z+w)= e^{-{1\over 2}(z \bar w-\bar z w)}D(w)D(z), \qquad z,\quad w\in  \mathbb{C}
\end{eqnarray}
we infer the well-known commutation relation
\begin{eqnarray}
D(z)D(w)= e^{(z \bar w-\bar z w)}D(w)D(z).
\end{eqnarray}

Coherent states were invented by Schr\"odinger in 1926 in  the context of the quantum harmonic oscillator. They were defined as minimum-uncertainty states that exhibit its classical behavior  \cite{Schrodinger}. 
In 1963, they have been simultaneous rediscovered by Glauber \cite{Glauber1,Glauber2}, Klauder \cite{Klauder1a,Klauder1b} and Sudarshan \cite{Sudarshan} in   quantum optics of coherent light beams emitted by lasers.
Since there, they became very popular objects in mathematics (specially in functional analysis, group theory and representations, geometric quantization, etc.), and in nearly all branches of quantum physics (nuclear,  atomic and solid state physics, statistical mechanics, quantum electrodynamics, path integral, quantum field theory, etc.). For more information
we refer the reader to the references \cite{Ali&al,Klauder&Skagerstam,Perelomov72,Zhang&al90}. 

The vast field covered  by coherent states motivated their generalizations to other families of states deducible from noncanonical operators and satisfying not necessarily all  above mentioned  properties.

The first  class of generalizations, based on the  equivalent conditions given in Definition \ref{CSdef1}, include:\\
a) The approach  by Barut and Girardello \cite{Barut}  considering coherent states as eigenstates of the annihilation operator. This approach was unsuccessful because of its drawbacks from both mathematical and physics point of view  as
detailed in  \cite{Gilmore74a,Perelomov72}.\\
b) The approach based on the  minimum-uncertainty states, i.e.  essentially on the original motivation of Schr\"odinger in his construction of wavepackets which follow the motion of a classical particle while retaining their shapes. This was the basis for building the intelligent coherent states for various dynamical systems \cite{Aragone&al76, Aragone&al74, Nieto&Simmons78,Nieto&Simmons79, Nieto&al81}. Nevertheless, as has been emphasized by Zhang {\it et al} \cite{Zhang&al90}, such a  generalization has several limitations.\\
c) The approach related to the unitary representation of the group generated by the creation and annihilation operators. In two papers by Klauder \cite{Klauder1a,Klauder1b} devoted to a set of continuous states, one finds the basic ideas of  coherent states construction for arbitrary Lie groups, which have been exploited by  Gilmore \cite{Gilmore} and  Perelomov \cite{Perelomov72, Perelomov}  to formulate a general and complete formalism of building coherent states for various deformations of the Heisenberg group with  properties similar to those of the harmonic oscillator. The key result of  this development was the  intimate connection of  the coherent states  with the dynamical group of  a given  physical system.

 Two other generalizations complete this first class of generalizations:  $(i)$  the covariant coherent states introduced in Ref. \cite{Ali&al},  considered as a generalization of Gilmore-Perelomov formalism  in the sense that the CS are built from more general groups (homogeneous spaces), and $(ii)$  the nonlinear coherent states  related to nonlinear algebras. Even though nonlinear coherent states have been used to analyze some quantum mechanical systems as  the motion of a trapped ion \cite{Junker&Roy,deMatos&Vogel}, they are not merely mathematical objects. They were defined as right eigenstates of a generalized annihilation operator \cite{Manko&al,deMatos&Vogel}.

The second class of generalizations is essentially based on the overcompleteness property of coherent states. This property was the {\it raison d'\^etre} of the mathematically oriented construction of generalized coherent states by Ali {\it et al} \cite{Ali&al99,Ali&al} or of the ones with physical orientations \cite{Gazeau09,Gazeau&Klauder,Klauder&al}. Numerous publications continue to appear  using this property, see for example \cite{Daoud,Hounkonnou&Ngompe,Hounkonnou&Sodoga,Popov} and references therein. The overcompleteness property is the most important criteria to be satisfied by CS  as required by  Klauder's criteria \cite{Klauder&al}. 

To end this quick overview, let us   mention  the generalization performed through the so-called {\it coherent state map,} elaborated by Odzijewicz \cite{Odzijewicz98} in 1998 and generalized in \cite{Hounkonnou&Bukweli10}.  It is now known that  the coherent state
map may be used as a tool for  the geometric quantization {\it \`a la Kostant-Souriau}  \cite{Odzijewicz98}. See the works by Kirillov \cite{Kirillov} and Kostant \cite{Kostant} for details on geometric quantization.

\begin{definition}
We  call deformed Heisenberg  algebra,  an associative algebra generated by the set of operators $\{\mathbf{1},\; a,\; a^\dagger,\; N\}$
satisfying the relations
\begin{eqnarray}\label{dal}
 [N,\;a^\dagger]= a^\dagger,\qquad [N,\;a]= -a,\label{uq01}
\end{eqnarray}
such that there exists a non-negative analytic function $\varphi$, called the structure function,  defining the operator products $ a^\dagger a$ and $aa^\dagger$ in the following way:
\begin{eqnarray}
 a^\dagger a := \varphi(N),\qquad aa^\dagger:= \varphi(N+\mathbf{1}),\label{uq02}
\end{eqnarray}
where $N$ is a self-adjoint operator, $a$ and its Hermitian conjugate $a^\dagger$ denote the deformed annihilation and creation operators, respectively.

\end{definition}
The function  $\varphi$,  encoding all required information,  for instance, in the construction of irreducible representations of the algebra,   remains the main task to solve when one deals with  such deformed algebra (\ref{uq01}). Different approaches for its determination  are spread in the literature. See  \cite{Baloitcha,Borzov,Burban1,Kosinski,Meljanac} and references therein.
More importantly, as it will be shown in the sequel, the structure function will be the key for  the  unification of  the coherent state construction methods  from generalized algebras, respecting    Klauder criteria. 
Note that the method put forward by Klauder \cite{Klauder&al} is based on  an appropriate  choice of a set of strictly positive parameters. In the present paper,  such a  set of positive parameters is determined by the  structure function.

The paper is  organized as follows. In Section \ref{Sec2}, the deformed Heisenberg algebra is described and  the   structure function is deduced. The spectrum of  the associated deformed oscillator is computed. The Section \ref{Sec3} is devoted to the construction of the deformed coherent states using the Klauder approach.  In section \ref{Sec4},  quantum statistics and geometry  of the deformed coherent states are investigated. Concluding remarks end the paper in Section \ref{Sec5}.

\section{$(q;l,\lambda)$-deformed Heisenberg algebra} \label{Sec2}
 Consider now  the following $(q;l,\lambda)$-deformed Heisenberg algebra  generated by operators $N$, $a$, $a^\dagger$
satisfying 
\begin{eqnarray}
[N,\;a]=-a,\qquad [N,\;a^\dagger]= a^\dagger,\label{Kalnins1}
\end{eqnarray}
with the operator products 
\begin{eqnarray}
&& aa^\dagger-a^\dagger a =l^2q^{-N+\lambda-1}.\label{Kalnins2}
\end{eqnarray}
One can readily check that the commutator $[.,\;.]$ of operators is antisymmetric and satisfies the Jacobi identity conferring  a Lie algebra structure to the $(q;l,\lambda)$-deformed Heisenberg algebra. This algebra plays an important role in  mathematical sciences in general, and, in  particular, in mathematical physics. In a notable work \cite{Kalnins},
similar associative algebra has been investigated  by Kalnins {\it et al}
 under the form:
\begin{eqnarray}
&&[H,\;E_+]= E_+\qquad [H,\;E_-]=-E_-\cr
&&[E_+,\;E_-]=-q^{-H}\mathcal{E}\qquad [\mathcal{E},\;E_{\pm}]=0=[\mathcal{E},\;H],
\end{eqnarray}
where $q$ is a  real number such that $0<q<1$. These authors showed that
the elements $\mathcal{C}= qq^{-H}\mathcal{E}+(q-1)E_+E_-$ and $\mathcal{E}$
lie in the center of this algebra.  It admits a class of irreducible representations for $\mathcal{C}=l^2I$ and $\mathcal{E}= l^2q^{\lambda-1}I$, where $l$ and $\lambda$ are real numbers with $l\neq0$.

The $(q;l,\lambda)$-deformed Heisenberg algebra (\ref{Kalnins1}) is a generalized algebra in the sense that it can generate a series of existing algebras as particular cases. For instance, even
 the generalization of the Quesne-algebra performed in \cite{Hounkonnou&Ngompe,Quesne} can  be deduced from (\ref{Kalnins1}) by setting   $l=1$ and $\lambda=0$.

In the sequel, we consider the Fock space of the Bose oscillator constructed as follows. From the vacuum vector $|0\rangle$ defined by $a|0\rangle=0$, the normalized vectors $|n\rangle$ for $n\ge 1,$ i.e. eigenvectors of the operator $N,$ are obtained as $|n\rangle=C_n(a^\dagger)^n|0\rangle$, where $C_n$ stands for some  normalization constant to be determined. 

\begin{proposition}
The structure function of the $(q; l, \lambda)$-deformed Heisenberg algebra $(\ref{Kalnins1})-(\ref{Kalnins2})$ is given by
\begin{eqnarray}\label{strucure}
\varphi(n)= l^2q^\lambda\frac{1-q^{-n}}{q-1}=l^2q^{\lambda-n}[n]_q,\;\; q>0,
\end{eqnarray}
where $[n]_q= \frac{1-q^n}{1-q},$ with $0<q<1$ or $1<q,$ is the $q_n-$ factors (also known as $q$-deformed numbers in Physics literature  \cite{Gasper}).
\end{proposition}
{\bf Proof:} From the definition (\ref{uq02}),  $a^\dagger a=\varphi(N)$ and $aa^\dagger= \varphi(N+1).$ Thus, (\ref{Kalnins2}) is written as
\begin{eqnarray*}
 \varphi(N+\mathbf{1})-\varphi(N)=l^2q^{-N+\lambda-1}.
\end{eqnarray*}
Applying this relation to the vectors $|n\rangle,$  we obtain the recurrence relation 
\begin{eqnarray*}
\varphi(n+1)-\varphi(n)=l^2q^{\lambda-n-1},\quad \forall n\in\mathbb{N}
\end{eqnarray*}
from which we deduce
\begin{eqnarray*}
 \varphi(n)= \varphi(0)+l^2q^\lambda\frac{1-q^{-n}}{q-1}.
\end{eqnarray*}
Since, in particular, $\varphi(N)|0\rangle=a^\dagger a|0\rangle=0$ implies $\varphi(0)|0\rangle=0$, we have $\varphi(0)=0.$ Then (\ref{strucure}) follows.
The structure  function  is also a strictly increasing function for $x\in\mathbb{R}$ since  
\begin{eqnarray*}
 \frac{d\varphi(x)}{dx}= l^2q^{\lambda-x}\frac{\ln q}{q-1}>0,\;\mbox{ for } q>0.
\end{eqnarray*}
Since $\varphi(0)=0$, it follows that $\varphi(x)\geq0$ for any real $x>0$ and in particular $\varphi(n)\geq 0$, $\forall n\geq0$. \hfill$\square$

\begin{proposition}\label{corola1}
The orthonormalized basis of the Fock space $\mathcal{F}$ is given by
\begin{eqnarray}\label{fockstate}
 |n\rangle= \frac{q^{n(n+1)/4}}{\sqrt{(l^2q^\lambda)^n[n]_q!}}(a^\dagger)^n|0\rangle,\qquad n=0,\;1,\;2,\; ...
\end{eqnarray}
where $[0]_q!:=1$ and $[n]_q!:=[n]_q[n-1]_q...[1]_q$.\\
Moreover, the action of the operators $a$, $a^\dagger$, $N$, $a^\dagger a$ and $aa^\dagger$ on the  vectors $|n\rangle$ for $n\ge 1$ are given by 
\begin{eqnarray}
&&a|n\rangle=\sqrt{l^2q^{\lambda-n}[n]_q}|n-1\rangle,\\&& 
a^\dagger|n\rangle=\sqrt{l^2q^{\lambda-n-1}[n+1]_q}|n+1\rangle,\\&& 
N|n\rangle=n|n\rangle, \\&&  a^\dagger a|n\rangle=l^2q^{\lambda-n}[n]_q|n\rangle,\\&& 
aa^\dagger|n\rangle=l^2q^{\lambda-n-1}[n+1]_q|n\rangle.
\end{eqnarray}
\end{proposition}
{\bf Proof:} To determine the constant of normalization $C_n,$ we set
\begin{eqnarray*}
 1=:\langle n|n\rangle=|C_n|^2\langle 0|a^n(a^\dagger)^n|0\rangle= |C_n|^2\varphi(n)\varphi(n-1)...\varphi(1)\langle 0|0\rangle
\end{eqnarray*}
leading to $C_n=\frac{q^{n(n+1)/4}}{\sqrt{(l^2q^\lambda)^n[n]_q!}}$. Replacing $C_n$ by its value in the definition of $|n\rangle$ given above yields (\ref{fockstate}). The orthogonality of the vectors $|n\rangle$ is a direct consequence of $a|0\rangle=0$. The rest of the proof is obtained from (\ref{fockstate}) using (\ref{Kalnins1}), (\ref{Kalnins2}) and (\ref{strucure}).\hfill$\square$

\begin{theorem}\label{thm}
 The operators $(a+a^\dagger)$ and $i(a-a^\dagger),$ defined on the Fock space $\mathcal{F},$ are bounded and, consequently, self-adjoint  if $q>1$. If $q<1$, they are not self-adjoint. 
\end{theorem}
{\bf Proof:} The matrix elements of  the operator $(a+a^\dagger)$ on the basis $|n\rangle$ are given by
\begin{eqnarray}\label{jacobi01}
\langle m|(a+a^\dagger)|n\rangle=x_n\delta_{m, n-1}+x_{n+1}\delta_{m, n+1}, n,\;m=0,\;1,\;2,\;\cdots 
\end{eqnarray}
while the matrix elements of the operator  $i(a-a^\dagger)$  are given by
\begin{eqnarray}\label{jacobi02}
\langle m|i(a-a^\dagger)|n\rangle=ix_n\delta_{m, n-1}-ix_{n+1}\delta_{m, n+1}, n,\;m=0,\;1,\;2,\;\cdots
\end{eqnarray}
where $x_n= \left(l^2q^{\lambda-n}[n]q\right)^{1/2}$.
Besides, the operators $(a+a^\dagger)$ and $i(a-a^\dagger)$ can be represented by the two following symmetric Jacobi matrices, respectively:
\begin{eqnarray}\label{jacobir}
 \left(\begin{array}{cccccc}0&x_1&0&0&0&\cdots\\x_1&0&x_2&0&0&\cdots\\0&x_2&0&x_3&0&\cdots\\\vdots&\ddots&\ddots&\ddots&\ddots&\ddots
       \end{array}\right)
\end{eqnarray}
and
\begin{eqnarray}\label{jacobic}
 \left(\begin{array}{cccccc}0&-ix_1&0&0&0&\cdots\\ix_1&0&-ix_2&0&0&\cdots\\0&ix_2&0&-ix_3&0&\cdots\\\vdots&\ddots&\ddots&\ddots&\ddots&\ddots
       \end{array}\right)
\end{eqnarray}
Two situations deserve investigation:
\newline
$\bullet$ Suppose    $q>1$. Then,
\begin{eqnarray*}
 \left|x_n\right|=\left(\frac{l^2q^\lambda}{q-1}\frac{q^n-1}{q^n}\right)^{1/2}<\left(\frac{l^2q^\lambda}{q-1}\right)^{1/2},\; \forall n\geq1.
\end{eqnarray*}
Therefore, the Jacobi matrices  in (\ref{jacobir}) and (\ref{jacobic}) are bounded and self-adjoint (Theorem 1.2., Chapter VII in Ref. \cite{Berezanskii}). Thus, $(a+a^\dagger)$ and $i(a-a^\dagger)$ are bounded and, consequently, self-adjoint.\\
$\bullet$ Contrarily, if $q<1,$ then
\begin{eqnarray}
 \lim_{n\to\infty}x_n=\lim_{n\to\infty}\left(l^2q^\lambda\frac{1-q^{-n}}{q-1}\right)^{1/2}=\infty.
\end{eqnarray}
Considering the series $\sum_{n=1}^\infty 1/x_n$, we obtain 
\begin{eqnarray*}
\overline{\lim_{n\to\infty}}\left(\frac{1/x_{n+1}}{1/x_n}\right)=\overline{\lim_{n\to\infty}}\left(\frac{1-q^{-n}}{1-q^{-n-1}}\right)^{1/2}= q^{1/2}<1.
\end{eqnarray*}
This ratio test  leads to the conclusion  that the series $\sum_{n=1}^\infty 1/x_n$ converges. Moreover, $1-2q+q^2=(1-q)^2\geq0\Longrightarrow q^{-1}+q\geq2$. Hence,
\begin{eqnarray*}
&& 0\leq\left(\frac{l^2q^\lambda}{q-1}\right)^2\left(1-q^{-n}(q+q^{-1})+q^{-2n}\right)\leq\left(1-2q^{-n}+q^{-2n}\right)\left(\frac{l^2q^\lambda}{q-1}\right)^2 \cr
\Leftrightarrow&&0\leq\left(l^2q^\lambda\frac{1-q^{-n+1}}{q-1}\right)\left(l^2q^\lambda\frac{1-q^{-n-1}}{q-1}\right)\leq \left(l^2q^\lambda\frac{1-q^{-n}}{q-1}\right)^2 \cr
\Leftrightarrow&&0\leq\left(l^2q^\lambda\frac{1-q^{-n+1}}{q-1}\right)^{1/2}\left(l^2q^\lambda\frac{1-q^{-n-1}}{q-1}\right)^{1/2}\leq \left(l^2q^\lambda\frac{1-q^{-n}}{q-1}\right) \cr
\Leftrightarrow &&
0\leq x_{n-1}x_{n+1}\leq x_n^2.
\end{eqnarray*}
Therefore, the Jacobi matrices  in (\ref{jacobir}) and (\ref{jacobic}) are not self-adjoint (Theorem 1.5., Chapter VII in Ref. \cite{Berezanskii}).\hfill$\square$

\begin{definition}
The $(q;l,\lambda)$-deformed position, momentum  
and Hamiltonian  operators denoted by $X_{l,\lambda,q}$, $P_{l,\lambda,q}$ and $H_{l,\lambda,q},$ respectively, are defined as follows:
\begin{eqnarray}
 X_{l,\lambda,q}&:=& \left({\hbar}/{2\mathbf{ m}\omega}\right)^{1/2}(a+a^\dagger),\cr
P_{l,\lambda,q}&:=& -i\left({\mathbf{ m}\hbar\omega}/{2}\right)^{1/2}(a-a^\dagger)\cr
 H_{l,\lambda,q}&:=& \frac{1}{2\mathbf{ m}}(P_{l,\lambda,q})^2+\frac{1}{2}\mathbf{ m}\omega^2(X_{l,\lambda,q})^2
\cr&=& \frac{\hbar\omega}{2}(a^\dagger a+aa^\dagger).
\end{eqnarray}
\end{definition}

\begin{proposition} The following system characterization holds:
\begin{itemize}
\item The vectors $|n\rangle$ are eigenvectors of the $(q;l,\lambda)$-deformed Hamiltonian with respect to the eigenvalues
\begin{eqnarray}\label{mecaprop1}
 E_{l,\lambda,q}(n)= \frac{\hbar\omega}{2}l^2q^{\lambda-n-1}\big(q[n]_q+ [n+1]_q\big).
\end{eqnarray}
\item The mean values of $X_{l,\lambda,q}$ and $P_{l,\lambda,q}$ in the states $|n\rangle$ are zero while their variances are given by
\begin{eqnarray}
 (\Delta X_{l,\lambda,q})_n^2 &=& \frac{\mathbf{ m}\hbar\omega}{2}l^2q^{\lambda-n-1}\big(q[n]_q+ [n+1]_q\big),\label{mecaprop2}\\
(\Delta P_{l,\lambda,q})_n^2 &=& \frac{\hbar}{2\mathbf{ m}\omega}l^2q^{\lambda-n-1}\big(q[n]_q+ [n+1]_q\big),\label{mecaprop3}
\end{eqnarray}
where $(\Delta A)_n^2=\langle A^2\rangle_n-\langle A\rangle_n^2$  with $\langle A\rangle_n=\langle n|A|n\rangle$.

\item The position-momentum uncertainty relation is given by
\begin{eqnarray}\label{mecaprop4}
 (\Delta X_{l,\lambda,q})_n(\Delta P_{l,\lambda,q})_n= \frac{h}{2}l^2q^{\lambda-n-1}\big(q[n]_q+ [n+1]_q\big)
\end{eqnarray}
which is reduced, for the vacuum state, to the expression
\begin{eqnarray}\label{mecaprop5}
 (\Delta X_{l,\lambda,q})_0(\Delta P_{l,\lambda,q})_0= \frac{h}{2}l^2q^{\lambda-1}.
\end{eqnarray}
\end{itemize}
\end{proposition}
{\bf Proof:} Indeed, using the result of the Proposition \ref{corola1}, we get
\begin{eqnarray*}
 H_{l,\lambda,q}|n\rangle= \frac{\hbar\omega}{2}(a^\dagger a+aa^\dagger)\rangle= \frac{\hbar\omega}{2}l^2q^{\lambda-n-1}\big(q[n]_q+ [n+1]_q\big)|n\rangle.
\end{eqnarray*}
The  first two relations (\ref{jacobi01}) and (\ref{jacobi02}) in the proof of the previous Theorem \ref{thm} yield
$\langle n|(a+a^\dagger)|n\rangle=0=\langle n|i(a-a^\dagger)|n\rangle$ and 
$\langle n|(a+a^\dagger)^2|n\rangle=x_n^2+x_{n+1}^2=\langle n|i^2(a-a^\dagger)^2|n\rangle$. Therefore, $\langle n|X_{l,\lambda}|n\rangle=0=\langle n|P_{l,\lambda}|n\rangle$,
$\langle n|X^2_{l,\lambda}|n\rangle=\frac{\mathbf{ m}\hbar\omega}{2}(x_n^2+x_{n+1}^2)$ and $\langle n|P^2_{l,\lambda}|n\rangle=\frac{\hbar}{2\mathbf{ m}\omega}(x_n^2+x_{n+1}^2)$. The rest of the proof is obtained replacing $x_n$ and  $x_{n+1}$ by their expressions.\hfill$\square$

\section{Coherent states $|z\rangle_{l,\lambda}$}\label{Sec3}

\begin{definition}
The coherent states associated with the algebra (\ref{Kalnins1})-(\ref{Kalnins2}) are defined as
\begin{eqnarray}\label{KalninCS}
 |z\rangle_{l,\lambda}:= \mathcal{N}_{l,\lambda}^{-1/2}(|z|^2)\sum_{n=0}^\infty\frac{q^{n(n+1)/4}z^n}{\sqrt{(l^2q^\lambda)^n[n]_q!}}|n\rangle,\; z\in\mathbf{D}_{l,\lambda},
\end{eqnarray}
where
\begin{eqnarray}
 \mathcal{N}_{l,\lambda}(x)&=&\sum_{n=0}^\infty\frac{q^{n(n+1)/2}x^n}{(l^2q^\lambda)^n[n]_q!}
=\sum_{n=0}^\infty\frac{q^{n(n-1)/2}}{(q;q)_n}\left(\frac{(1-q)qx}{ l^2q^\lambda}\right)^n
\end{eqnarray}
and
\begin{eqnarray}
 \mathbf{D}_{l,\lambda}=\left\{z\in\mathbb{C}:\; |z|^2<R_{l,\lambda}\right\},\;
\mbox{with }\; R_{l,\lambda}=\left\{\begin{array}{lcl} \infty&\mbox{if }& 0<q<1\\\frac{l^2q^\lambda}{q-1}&\mbox{if }& q>1.\end{array}\right.
\end{eqnarray}
\end{definition}
 $R_{l,\lambda}$ is the   convergence radius of the series $\mathcal{N}_{l,\lambda}(x)$.

Remark  that the $q$-deformed coherent states introduced in \cite{Quesne} are recovered as a particular case  corresponding to $l=1$ and $\lambda=0$.

We now aim  at showing that  the coherent states (\ref{KalninCS}) satisfy the Klauder's criteria \cite{Klauder&Skagerstam,Klauder&al}. To this end let us first prove the following lemma:

\begin{lemma}\label{Buklemma} If $q>1,$ then
\begin{eqnarray}
&& \frac{\mathcal{N}_{l,\lambda}(x)}{\mathcal{N}_{l,\lambda}(q^{-1}x)}=\frac{1}{1-(q-1)x/(l^2q^\lambda)}\;,\label{Kaprop1}\\
&&\mathcal{N}_{l,\lambda}(x)= \frac{1}{\left((q-1)x/(l^2q^\lambda);q^{-1}\right)_\infty},\label{Kaprop2}\\
&&\int_0^{R_{l,\lambda}}x^n\left(\mathcal{N}_{l,\lambda}(q^{-1}x)\right)^{-1}d_q^{l,\lambda}x= (l^2q^\lambda)^nq^{-n(n+1)/2}[n]_q!.\label{Kaprop3}
\end{eqnarray}
\end{lemma}
{\bf Proof:} 
We use  the $(q;l,\lambda)$-derivative defined by
\begin{eqnarray}\label{Kaderiva}
 \partial_q^{l,\lambda}f(x)= l^2q^\lambda\frac{f(x)-f(q^{-1}x)}{(q-1)x}
\end{eqnarray}
to obtain
\begin{eqnarray*}
\mathcal{N}_{l,\lambda}(x)=\partial_q^{l,\lambda}\mathcal{N}_{l,\lambda}(x)=
l^2q^\lambda\frac{\mathcal{N}_{l,\lambda}(x)-\mathcal{N}_{l,\lambda}(q^{-1}x)}{(q-1)x}
\end{eqnarray*}
which leads to (\ref{Kaprop1}) and 
\begin{eqnarray}
\mathcal{N}_{l,\lambda}(x)=\frac{\mathcal{N}_{l,\lambda}(q^{-n}x)}{\prod_{k=0}^{n-1}\big(1-(q-1)q^{-k}x/(l^2q^\lambda)\big)}, \; n=1,\;2,\;...
\end{eqnarray}
Letting $n$ to $+\infty$ and taking into account the fact that $\mathcal{N}_{l,\lambda}(0)=1$ lead to (\ref{Kaprop2}).

Next, we use  the $(q;l,\lambda)$-integration given by
\begin{eqnarray}\label{Kaintegra}
\int_0^af(x)d_q^{l,\lambda}x=\frac{q-1}{l^2q^\lambda}a\sum_{k=0}^\infty q^{-k}f(aq^{-k})
\end{eqnarray}
to get
\begin{eqnarray*}
\int_0^{R_{l,\lambda}}x^n\left(\mathcal{N}_{l,\lambda}(q^{-1}x)\right)^{-1}d_q^{l,\lambda}x &=& \sum_{k=0}^\infty q^{-(n+1)k}\frac{(l^2q^\lambda)^n}{(q-1)^n}\big(q^{-(k+1)};q^{-1}\big)_\infty
\cr&=&\frac{(l^2q^\lambda)^n}{(q-1)^n}\big(q^{-1};q^{-1}\big)_\infty
\sum_{k=0}^\infty \frac{q^{-(n+1)k}}{\big(q^{-1};q^{-1}\big)_k}
\cr&=&\frac{(l^2q^\lambda)^n}{(q-1)^n}\frac{\big(q^{-1};q^{-1}\big)_\infty}{\big(q^{-(n+1)};q^{-1}\big)_\infty}
\cr&=&\frac{(l^2q^\lambda)^n}{(q-1)^n}\big(q^{-1};q^{-1}\big)_n
=(l^2q^\lambda)^nq^{-n(n+1)/2}[n]_q!.
\end{eqnarray*}
\hfill$\Box$

\begin{proposition}
The coherent states defined in (\ref{KalninCS})
\begin{itemize}
 \item [(i)] are normalized eigenvectors of the operator $a$ with eigenvalue $z$, i.e.
\begin{eqnarray}
a|z\rangle_{l,\lambda}=z|z\rangle_{l,\lambda},\qquad {}_{l,\lambda}\langle z|z\rangle_{l,\lambda}=1;
\end{eqnarray}
\item[(ii)] are not orthogonal to each other, i.e.
\begin{eqnarray}
 {}_{l,\lambda}\langle z_1|z_2\rangle_{l,\lambda} \neq0,\;\;\mbox{when}\;\;z_1\neq z_2;
\end{eqnarray}
\item[(iii)]are continuous in their labels $z$;
\item[(iv)] resolve the unity, i.e.
\begin{eqnarray}\label{unitf}
 \mathbf{1}=\int_{\mathbf{D}_{l,\lambda}} d\mu_{l,\lambda}(\bar z,z)|z\rangle_{l,\lambda} {}_{l,\lambda}\langle z|,
\end{eqnarray}
where
\begin{eqnarray}
 d\mu_{l,\lambda}(\bar z,z)=\frac{1-q}{l^2q^\lambda\ln{q^{-1}}}\frac{\mathcal{N}_{l,\lambda}(\bar zz)}{\mathcal{N}_{l,\lambda}(\bar zz/q)}\frac{d^2z}{\pi},\;\mbox{ if }\; 0<q<1,
\end{eqnarray}
and
\begin{eqnarray}
 d\mu(\bar z,z)=\frac{1}{2\pi}\frac{d_q^{l,\lambda}x\;d\theta}{1-(q-1)x/(l^2q^\lambda)}, \quad x=|z|^2,\;\;\theta=\arg(z),
\end{eqnarray}
with $0<x<\frac{l^2q^\lambda}{q-1}$ and $0\leq\theta\leq2\pi$ for $q>1$.
\end{itemize}
\end{proposition}
{\bf Proof:}
\\
{$\bullet$\it Non orthogonality and normalizability}
\begin{eqnarray}
 {}_{l,\lambda}\langle z_1|z_2\rangle_{l,\lambda}= \frac{\mathcal{N}_{l,\lambda}(\bar z_1z_2)}{\left(\mathcal{N}_{l,\lambda}(|z_1|^2)\mathcal{N}_{l,\lambda}(|z_2|^2)\right)^{1/2}}\neq 0
\end{eqnarray}
imply that the coherent states are not orthogonal.
\\
{$\bullet$\it Normalizability}
\\
From the above relation taking $z_1=z_2=z$ we obtain ${}_{l,\lambda}\langle z|z\rangle_{l,\lambda}=1$. Also,
\begin{eqnarray*}
a|z\rangle_{l,\lambda}&=&\mathcal{N}_{l,\lambda}^{-1/2}(|z|^2)\sum_{n=0}^\infty\frac{q^{n(n+1)/4}z^n}{\sqrt{(l^2q^\lambda)^n[n]_q!}}a|n\rangle\cr
&=&\mathcal{N}_{l,\lambda}^{-1/2}(|z|^2)\sum_{n=1}^\infty\frac{q^{n(n-1)/4}z^n}{\sqrt{(l^2q^\lambda)^{n-1}[n-1]_q!}}|n-1\rangle\cr
&=&z\mathcal{N}_{l,\lambda}^{-1/2}(|z|^2)\sum_{n=0}^\infty\frac{q^{n(n+1)/4}z^{n}}{\sqrt{(l^2q^\lambda)^n[n]_q!}}|n\rangle.
\end{eqnarray*}
\\
{$\bullet$\it Continuity in  the labels $z$}
\begin{eqnarray*}
 |||z_1\rangle_{l,\lambda}-|z_2\rangle_{l,\lambda}||^2= 2\left(1-\mathcal{R}e{}_{l,\lambda}\langle z_1|z_2\rangle_{l,\lambda}\right).
\end{eqnarray*}
So, $|||z_1\rangle_{l,\lambda}-|z_2\rangle_{l,\lambda}||^2\to 0$ as $|z_1-z_2|\to 0$, since ${}_{l,\lambda}\langle z_1|z_2\rangle_{l,\lambda}\to 1$ as $|z_1-z_2|\to 0$.
\\
{$\bullet$\it Resolution of the unity}

The computation of the RHS of (\ref{unitf}) gives
\begin{eqnarray}
 \int_{\mathbf{D}_{l,\lambda}} d\mu_{l,\lambda}(\bar z,z)|z\rangle_{l,\lambda}{}_{l,\lambda}\langle z|= \sum_{n,m}|n\rangle\langle m|\frac{q^{[n(n+1)+m(m+1)]/4}}{\sqrt{(l^2q^\lambda)^{n+m}[n]_q![m]_q!}}
\int_{\mathbf{D}_{l,\lambda}} \bar z^nz^m\frac{d\mu_{l,\lambda}(\bar z,z)}{\mathcal{N}_{l,\lambda}(|z|^2)}.
\end{eqnarray}
So, in order to satisfy (\ref{unitf}) it is required 
\begin{eqnarray}
\int_{\mathbf{D}_{l,\lambda}} \bar z^nz^m\frac{d\mu_{l,\lambda}(\bar z,z)}{\mathcal{N}_{l,\lambda}(|z|^2)}=\delta_{m n}(l^2q^\lambda)^n q^{-n(n+1)/2}[n]_q!,\quad n,\;m= 0,\; 1,\; 2,\; ...
\end{eqnarray}
Upon passing to polar coordinates,  $z=\sqrt x\;e^{i\theta}$, $d\mu_{l,\lambda}(\bar z,z)= d\omega_{l,\lambda}(x)d\theta$
 where $0\leq\theta\leq2\pi$, $0<x< R_{l,\lambda}$  and $\omega_{l,\lambda}$ is a positive valued function, this is equivalent to 
the classical Stieltjes power moment problem when $0<q<1$ or the Hausdorff power moment problem when $q>1$ \cite{Akhiezer, Tarmakin}:
\begin{eqnarray}\label{KaMoment}
 \int_0^{R_{l,\lambda}}x^n\;\frac{2\pi\;d\omega_{l,\lambda}(x)}{\mathcal{N}_{l,\lambda}(x)}=(l^2q^\lambda)^n q^{-n(n+1)/2}[n]_q!,\quad n= 0,\; 1,\; 2,\; ...
\end{eqnarray}

If $0<q<1,$ then we have the following Stieltjes power moment problem:
\begin{eqnarray}
 \int_0^{+\infty}x^n\frac{2\pi\;d\omega_{l,\lambda}(x)}{\mathcal{N}_{l,\lambda}(x)}= (l^2q^\lambda)^n q^{-n(n+1)/2}[n]_q!,
\end{eqnarray}
or, equivalently, 
\begin{eqnarray}
 \int_0^{+\infty}y^n\frac{2\pi\;d\omega_{l,\lambda}(l^2q^\lambda y)}{E_q\big((1-q)qy\big)}
=q^{-n(n+1)/2}[n]_q!, 
\end{eqnarray}
where the change of variable $y=\frac{x}{l^2q^\lambda}$ has been made.
Atakishiyev and Atakishiyeva \cite{Atakishiyeva} have proved that
\begin{eqnarray}
 g_q(n)= \int_0^{+\infty}\frac{y^{n-1}dy}{E_q((1-q)y)}= \frac{\ln{q^{-1}}}{1-q}q^{-n(n-1)/2}[n-1]!_q.
\end{eqnarray}
Therefore we deduce
\begin{eqnarray*}
 d\omega_{l,\lambda}(l^2q^\lambda y)=\frac{1}{2\pi}\frac{1-q}{\ln{q^{-1}}}\frac{E_q((1-q)qy)dy}{E_q((1-q)y)}
\end{eqnarray*}
or
\begin{eqnarray}
 d\omega_{l,\lambda}(x)&=&\frac{1}{2\pi}\frac{1-q}{l^2q^\lambda\ln{q^{-1}}}\frac{E_q((1-q)qx/(l^2q^\lambda))dx}{E_q((1-q)x/(l^2q^\lambda ))}
\cr&=&\frac{1}{2\pi}\frac{1-q}{l^2q^\lambda\ln{q^{-1}}}\frac{\mathcal{N}_{l,\lambda}(x)dx}{\mathcal{N}_{l,\lambda}(x/q)}.
\end{eqnarray}
Hence
\begin{eqnarray}
 d\mu_{l,\lambda}(\bar z,z)=\frac{1-q}{l^2q^\lambda\ln{q^{-1}}}\frac{\mathcal{N}_{l,\lambda}(\bar zz)}{\mathcal{N}_{l,\lambda}(\bar zz/q)}\frac{d^2z}{\pi}.
\end{eqnarray}

In the opposite, if $q>1,$ then combining (\ref{KaMoment}), (\ref{Kaprop1}) and (\ref{Kaprop2}) of the Lemma \ref{Buklemma} we get
\begin{eqnarray}
 d\mu(\bar z,z)=\frac{1}{2\pi}\frac{d_q^{l,\lambda}x\;d\theta}{1-(q-1)x/(l^2q^\lambda)}, \quad x=|z|^2,\;\;\theta=\arg(z),
\end{eqnarray}
where $0<x<\frac{l^2q^\lambda}{q-1}$ and $0\leq\theta\leq2\pi$.\hfill$\square$

\section{Statistics and geometry of coherent states $|z\rangle_{l,\lambda}$}\label{Sec4}

The conventional boson operators $b$ and $b^\dagger$ may be expressed in terms of the deformed operators $a$ and $a^\dagger$ as
\begin{eqnarray}
 b= a\;\sqrt{\frac{N}{\varphi(N)}}\quad\mbox{and}\quad b^\dagger=\sqrt{\frac{N}{\varphi(N)}}\;a^\dagger,\quad {\varphi(N)}\neq {\varphi(0)}
\end{eqnarray}
and their actions  on the states $|n\rangle$ are given by
\begin{eqnarray}
 b|n\rangle=\sqrt{n}|n-1\rangle,\quad \mbox{and} \quad 
b^\dagger|n\rangle=\sqrt{n+1}|n+1\rangle.
\end{eqnarray}
Besides,
\begin{eqnarray}
 b^r|n\rangle=\sqrt{\frac{n!}{(n-r)!}}|n-r\rangle, \qquad 0\leq r\leq n
\end{eqnarray}
and
\begin{eqnarray}
 (b^\dagger)^s|n\rangle=\sqrt{\frac{(n+s)!}{n!}}|n+s\rangle.
\end{eqnarray}
\subsection{\it  Quantum statistics of the coherent states $|z\rangle_{l,\lambda}$}
\begin{proposition}
 The expectation value of monomials  of boson creation and annihilation operators $b^\dagger$, $b$ in the coherent states $|z\rangle_{l,\lambda}$ are given by
\begin{eqnarray}
 \langle(b^\dagger)^sb^r\rangle
=\frac{\bar z^sz^r}{\mathcal{N}_{l,\lambda}(|z|^2)}\sum_{n=0}^\infty \sqrt{\frac{q^{[(n+s)(n+s+1)+(n+r)(n+r+1)]/2}(n+r)!(n+s)!}{(l^2q^\lambda)^{(n+s)+(n+r)}[n+s]_q![n+r]_q!}}\frac{|z|^{2n}}{n!},
\end{eqnarray}
where $s=0,\;1,\;2,\cdots$ and $r=0,\;1,\;2,\cdots$.\\
In particular,
\begin{eqnarray}
 \langle(b^\dagger)^rb^r\rangle= \frac{x^{r}}{\mathcal{N}_{l,\lambda}(x)}\left(\frac{d}{dx}\right)^r\mathcal{N}_{l,\lambda}(x),\quad x=|z|^2,\quad r=0,\;1,\;2,\cdots,
\end{eqnarray}
and
\begin{eqnarray}
 \langle N\rangle= x\frac{\mathcal{N}_{l,\lambda}'(x)}{\mathcal{N}_{l,\lambda}(x)}\;,
\end{eqnarray}
where $\mathcal{N}_{l,\lambda}'(x)$ denotes the derivative with respect to $x$.
\end{proposition}
{\bf Proof:} Indeed, for $s=0,\;1,\;2,\cdots$ and $r=0,\;1,\;2,\cdots$, we have 
\begin{eqnarray*}
&& \langle(b^\dagger)^sb^r\rangle:={}_{l,\lambda}\langle z|(b^\dagger)^sb^r|z\rangle_{l,\lambda}
\cr&&\quad=\frac{1}{\mathcal{N}_{l,\lambda}(|z|^2)}\sum_{m=0}^\infty\sum_{n=r}^\infty \sqrt{\frac{q^{[m(m+1)+n(n+1)]/2}n!(n-r+s)!}{(l^2q^\lambda)^{m+n}[m]_q![n]_q!(n-r)!(n-r)!}}\bar z^mz^n\langle m|n+s-r\rangle
\cr&&\quad=\frac{1}{\mathcal{N}_{l,\lambda}(|z|^2)}\sum_{n=r}^\infty \sqrt{\frac{q^{[(n+s-r)(n+s-r+1)+n(n+1)]/2}n!(n-r+s)!}{(l^2q^\lambda)^{n+s-r+n}[n+s-r]_q![n]_q!(n-r)!(n-r)!}}\bar z^{n+s-r}z^n
\cr&&\quad=\frac{\bar z^sz^r}{\mathcal{N}_{l,\lambda}(|z|^2)}\sum_{n=0}^\infty \sqrt{\frac{q^{[(n+s)(n+s+1)+(n+r)(n+r+1)]/2}(n+r)!(n+s)!}{(l^2q^\lambda)^{(n+s)+(n+r)}[n+s]_q![n+r]_q!}}\frac{|z|^{2n}}{n!},
\end{eqnarray*}

In the special case $s=r$, we have 
\begin{eqnarray*}
 \langle(b^\dagger)^rb^r\rangle&=& \frac{x^{r}}{\mathcal{N}_{l,\lambda}(x)}\sum_{n=0}^\infty \frac{q^{(n+r)(n+r+1)/2}(n+r)!}{(l^2q^\lambda)^{(n+r)}[n+r]_q!}\frac{x^{n}}{n!}
\cr&=& \frac{x^{r}}{\mathcal{N}_{l,\lambda}(x)}\sum_{n=r}^\infty \frac{q^{n(n+1)/2}(n)!}{(l^2q^\lambda)^{(n)}[n]_q!}\frac{x^{n-r}}{(n-r)!}
\cr&=& \frac{x^{r}}{\mathcal{N}_{l,\lambda}(x)}\left(\frac{d}{dx}\right)^r\mathcal{N}_{l,\lambda}(x),\quad x=|z|^2.
\end{eqnarray*}
In particular
\begin{eqnarray*}
 \langle N\rangle\equiv\langle b^\dagger b\rangle= x\frac{\mathcal{N}_{l,\lambda}'(x)}{\mathcal{N}_{l,\lambda}(x)}.
\end{eqnarray*}
\hfill$\square$

The probability of finding $n$ quanta in the deformed state $|z\rangle_{l,\lambda}$ is given by
\begin{eqnarray}
 \mathcal{P}_{l,\lambda}(n):=|\langle n|z\rangle_{l,\lambda}|^2= \frac{q^{n(n+1)/2}x^n}{(l^2q^\lambda)^n[n]_q!\mathcal{N}_{l,\lambda}(x)}.
\end{eqnarray}

The Mendel parameter  measuring the deviation from the Poisson statistics is defined by the quantity
\begin{eqnarray}
 Q_{l,\lambda}:=\frac{\langle N^2\rangle-\langle N\rangle^2-\langle N\rangle}{\langle N\rangle}.
\end{eqnarray}
Let us evaluate it explicitly.
From the expectation value of the operator  $N^2=(b^\dagger)^2b^2+ N$ provided by
\begin{eqnarray}
 \langle N^2\rangle
= x^2\frac{\mathcal{N}_{l,\lambda}''(x)}{\mathcal{N}_{l,\lambda}(x)}+
x\frac{\mathcal{N}_{l,\lambda}'(x)}{\mathcal{N}_{l,\lambda}(x)},
\end{eqnarray} 
we readily deduce
\begin{eqnarray}
 Q_{l,\lambda}=x\left(\frac{\mathcal{N}_{l,\lambda}''(x)}{\mathcal{N}_{l,\lambda}'(x)} -\frac{\mathcal{N}_{l,\lambda}'(x)}{\mathcal{N}_{l,\lambda}(x)}\right).
\end{eqnarray}
It is then worth noticing that for $x<<1$, 
\begin{eqnarray}
 Q_{l,\lambda}=-\frac{q(1-q)}{l^2q^\lambda(1+q)}x+ o(x^2)
\end{eqnarray}
 meaning that the $\mathcal{P}_{l,\lambda}(n)$ is a sub-Poissonian distribution \cite{Klauder&al}.

\subsection{\it  Geometry of the states $|z\rangle_{l,\lambda}$}
The geometry of a quantum state space can be described by the corresponding metric tensor. This real and positive definite metric is defined on the underlying manifold that the quantum states form, or belong to, by calculating the distance function (line element) between
two quantum states. So, it is also known as a Fubini-Study metric of the ray space. The knowledge of the quantum metric enables one to calculate quantum mechanical transition probability and uncertainties 

In the case $q<1$, the map from $z$ to $|z\rangle_{l,\lambda}$ defines a map from the space $\mathbb{C}$ of complex numbers onto
a continuous subset of unit vectors in Hilbert space and generates in the latter a two-dimensional surface with the following Fubini-Study metric:
\begin{eqnarray}
 d\sigma^2:= ||d|z\rangle_{l,\lambda}||^2-|_{l,\lambda}\langle z|d|z\rangle_{l,\lambda}|^2
\end{eqnarray}
\begin{proposition}
The above Fubini-Study metric  is reduced to
\begin{eqnarray}
 d\sigma^2= W_{l,\lambda}(x)d\bar z dz,
\end{eqnarray}
where $x=|z|^2$ and 
\begin{eqnarray}
 W_{l,\lambda}(x)=\left(x\frac{\mathcal{N}_{l,\lambda}'(x)}{\mathcal{N}_{l,\lambda}(x)}\right)'= \frac{d}{dx}\langle N\rangle.
\end{eqnarray}
In polar coordinates, $z= re^{i\theta}$,
\begin{eqnarray}
d\sigma^2= W_{l,\lambda}(r^2)(dr^2+r^2d\theta^2).
\end{eqnarray}
\end{proposition}
{\bf Proof:}
Computing $d|z\rangle_{l,\lambda}$ by taking into account the fact that any change of the form
$d|z\rangle_{l,\lambda}=\alpha|z\rangle_{l,\lambda}$, $\alpha\in\mathbb{C}$, has zero distance, we get
\begin{eqnarray*}
 d|z\rangle_{l,\lambda}= \mathcal{N}_{l,\lambda}(|z|^2)^{-1/2}\sum_{n=0}^\infty\frac{q^{n(n+1)/4}nz^{n-1}}{\sqrt{(l^2q^\lambda)^n[n]_q!}}|n\rangle\;dz.
\end{eqnarray*}
Then,
\begin{eqnarray*}
 ||d|z\rangle_{l,\lambda}||^2&=&\mathcal{N}_{l,\lambda}(|z|^2)^{-1}\sum_{n=0}^\infty\frac{q^{n(n+1)/2}n^2|z|^{2(n-1)}}{(l^2q^\lambda)^n[n]_q!}d\bar z dz
 \cr&=&\mathcal{N}_{l,\lambda}(|z|^2)^{-1}\left(\sum_{n=0}^\infty\frac{q^{n(n+1)/2}n|z|^{2(n-1)}}{(l^2q^\lambda)^n[n]_q!}\right.
\cr&&\quad\left.+|z|^2\sum_{n=0}^\infty\frac{q^{n(n+1)/2}n(n-1)|z|^{2(n-2)}}{(l^2q^\lambda)^n[n]_q!}\right)d\bar z dz
\cr&=&\mathcal{N}_{l,\lambda}(x)^{-1}\left(\mathcal{N}_{l,\lambda}'(x)+x\mathcal{N}_{l,\lambda}''(x)\right)d\bar z dz
\cr&=&\mathcal{N}_{l,\lambda}(x)^{-1}\left(x\mathcal{N}_{l,\lambda}'(x)\right)'d\bar z dz
\end{eqnarray*}
and
\begin{eqnarray*}
 |_{l,\lambda}\langle z|d|z\rangle_{l,\lambda}|^2&=&
\left|\mathcal{N}_{l,\lambda}(|z|^2)^{-1}\sum_{n=0}^\infty\frac{q^{n(n+1)/2}n|z|^{2(n-1)}}{(l^2q^\lambda)^n[n]_q!}\bar z dz\right|^2
\cr&=&x\mathcal{N}_{l,\lambda}(x)^{-2}\left(\mathcal{N}_{l,\lambda}'(x)\right)^2d\bar z dz.
\end{eqnarray*}
Therefore,
\begin{eqnarray*}
 d\sigma^2&=&\left(\mathcal{N}_{l,\lambda}(x)^{-1}\left(\mathcal{N}_{l,\lambda}'(x)+x\mathcal{N}_{l,\lambda}''(x)\right)-
x\mathcal{N}_{l,\lambda}(x)^{-2}\left(\mathcal{N}_{l,\lambda}'(x)\right)^2\right)d\bar{z}dz
\cr&=&\left(x\frac{\mathcal{N}_{l,\lambda}'(x)}{\mathcal{N}_{l,\lambda}(x)}\right)'d\bar{z}dz= \left(\frac{d}{dx}\langle N\rangle\right)d\bar{z}dz.
\end{eqnarray*}
\hfill$\square$

For $x<<1$, we have
\begin{eqnarray}
 W_{l,\lambda}(x)=\frac{q}{l^2q^\lambda}\left[1-\frac{2q(1-q)}{l^2q^\lambda(1+q)}x+o(x^2)\right].
\end{eqnarray}

 \section{Concluding remark}\label{Sec5}
In the present work, we have deformed the Heisenberg algebra   with the set of parameters
 $\{q, l,\lambda\}$  to generate  a new family of generalized 
coherent states respecting  the Klauder criteria. In this framework, 
 the matrix elements of  relevant operators have been  exactly computed and investigated from functional analysis point of view.
 Then, relevant statistical properties have been examined. Besides,
 a proof on the sub-Poissonian character of the statistics of  the main deformed states 
has been provided. This property   has been finally used to   determine   the induced generalized metric, characterizing 
the geometry of the considered system.

\section*{Acknowledgements}
This work is partially supported by the Abdus Salam International
Centre for Theoretical Physics (ICTP, Trieste, Italy) through the
Office of External Activities (OEA) - \mbox{Prj-15}. The ICMPA
is in partnership with
the Daniel Iagolnitzer Foundation (DIF), France.

\label{lastpage-01}
\end{document}